\documentclass[twocolumn,showpacs,showkeys,preprintnumbers,amsmath,amssymb]{revtex4}
\usepackage{graphicx}% Include figure files
\usepackage{dcolumn}% Align table columns on decimal point
\usepackage{bm}% bold math
\usepackage{docs}
\begin{document}
\title{Total cross sections for neutron-nucleus scattering}
\author{{S. V.   Suryanarayana$^1$\footnote{suryanarayan7@yahoo.com}}
, H. Naik$^2$, S. Ganesan$^3$, S. Kailas$^4$,  R. K. Choudhury$^1$  and  Guinyum Kim$^5$ }
%\author{{S. V.   SURYANARAYANA$^1$\footnote{suryanarayan7@yahoo.com}}
%, H. NAIK$^2$, S. GANESAN$^3$, S. KAILAS$^4$,  R. K. CHOUDHURY$^1$  AND  GUINYUM KIM$^5$ }
\affiliation{  $^1$  Nuclear  Physics Division, 
$^2$ Radio Chemistry Division, $^3$ Reactor Physics Design Division, $^4$ Director, Physics Group,\\
Bhabha Atomic Research Centre, Trombay, Mumbai 400 085, India, \\
$^5$ Department of Physics, Kyungpook National University, Deagu 702-701, Republic of Korea.}
%-------------------------------------------------------------------
%\documentclass[net-ND]{svjour}
% 
%\usepackage[tbtags]{amsmath}
%\usepackage{amssymb}
%\usepackage{graphicx}
%
%\usepackage{txfonts}
% Postscript TX Times-fonts

%\begin{document}

%\title{Neutron-nucleus scattering total cross sections}

%\author{ S.V. SURYANARAYANA \cormail{suryanarayan7@yahoo.com} \fnmsep 
%       \thanks{Present address: Nuclear Physics Division, Bhabha Atomic Research Centre, Mumbai 400 085, India}
%   \and S. GANESAN \inst{2}
%   \and S. KAILAS \inst{3}
%   \and R.K. CHOUDHURY \inst{1}
%   \and  KIM \inst{4} }
%}

%\institute{  Nuclear Physics Division,
%\and Reactor Physics Design Division,
%\and Director, Physics Group, Bhabha Atomic Research Centre, Mumbai 400085, INDIA,
%\and Korea Nuclear Data Center, Korea Atomic Energy Research Institute,
%            Deokjin-dong, Yuseong-gu, Daejeon, 305-353, Korea }
%}
%
%------------------------------------------------------------------

\begin{abstract}
%\abstract{
Systematics of neutron scattering cross sections  on  various materials for neutron energies up to several  hundred  MeV 
are  important for ADSS applications.  Ramsauer model is well known and widely applied to understand systematics of neutron nucleus 
total cross sections. In this work, we examined the role of  nuclear effective radius parameter (r$_0$) on Ramsauer model fits of neutron total cross sections.  
We  performed Ramsauer model global analysis of the experimental neutron total cross sections 
reported by W. P. Abfalterer, F. B. Bateman, {\it et. al.,},  from 20MeV to 550MeV  for nuclei ranging from Be to U .  
The global fit functions which can  fit total cross section data over periodic table are provided along with the required global  set of parameters.
The global fits predict within $\pm 8\%$ deviation to data, showing the scope for improvement.  
It has been observed that a finer adjustment of r$_0$ parameter alone can give very good  Ramsauer model description of 
neutron total scattering data within $\pm4\%$ deviation. The required r$_0$ values for Ramsauer model fits 
are shown as a function of nuclear mass number and an  empirical formula is suggested for r$_0$ values as a function of  mass number. 
In optical model approach for neutron scattering, we have modified the real part of 
Koning-Deleroche potentails to fit the neutron total cross sections using SCAT2 code.
The modified potentails  have a different energy dependence beyond 200MeV of neutron energy and fit the total 
cross sections from Al to Pb.

%}
\end{abstract}

\maketitle

\par
\begin{figure}
\includegraphics[width=9.00cm,height=10.0cm]{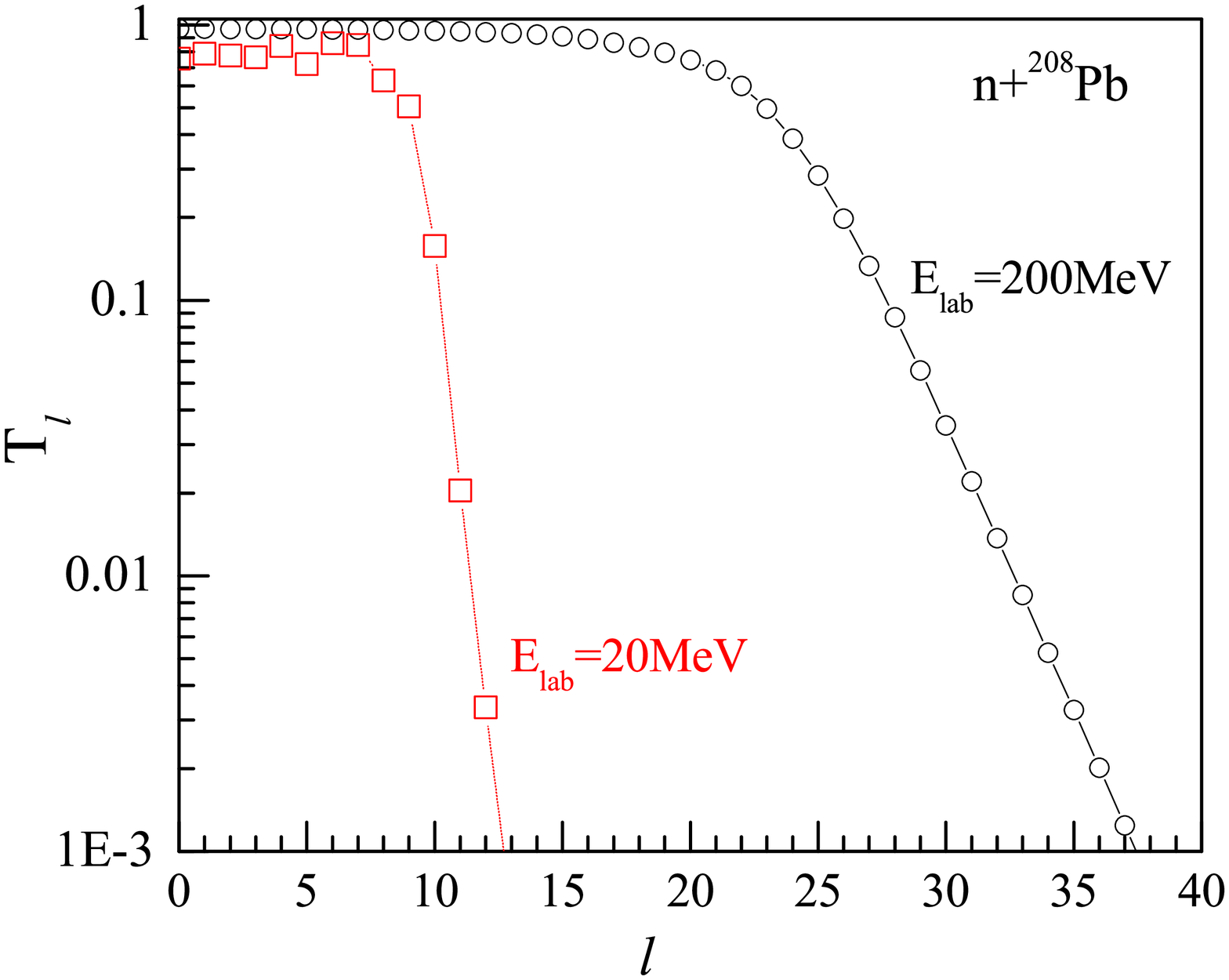}
\caption{Transmission function as a function of angular momentum for n+$^{208}$Pb system, showing the 
sharp cut off behaviour beyond some $\ell_{max}$. The results are generated by SCAT2 code using KD potentials. 
The neutron incident energies are indicated on curves.  Similar dependence is seen for all nuclei considered in Fig.\ref{sigtot}.}
\label{npbtl}
\end{figure}

\noindent
The concept of an accelerator driven sub-critical (ADS)
system is drawing   worldwide  attention  \cite{ads1,ads2}.  In  this   ADS
system,  neutrons  are produced by bombarding a heavy element target with a
high energy proton beam of typically above 1.0GeV with a current of $>10mA$
\cite{ads1}. Such a system serves a dual purpose of  energy  multiplication
and  waste  incineration. In this context it is important to study the systematics
of neutron scattering cross  sections  on  various  nuclei  for  neutron
energies up to  several  hundred  MeV.  In this
paper we propose an empirical approach to reproduce the  energy and atomic mass
dependence  of  total  cross sections ($\sigma_{tot}$).   The nuclear
Ramsauer model was first proposed by Lawson \cite{lawson} in the year 1953
as  a  simple means  to  understand  the  energy  dependence  of  total cross sections of
neutron nucleus scattering. In  order  to  appreciate  this  model,  it  is
necessary  to discuss briefly the optical model (OM) description of neutron
scattering. In the OM approach, the scattering amplitudes are obtained by 
solving the  Schrodinger's  equation  with complex optical model potentials (OMP), such as in 
SCAT2 optical model code \cite{scat2} with Koning and Deleroche potentials \cite{kd}.
The absorption arising from imaginary part of OMP gives the  reaction  cross
section.   The  calculations  are  usually  performed  using  partial  wave
expansion   method   and   the   phase    shifts $\delta_\ell$    
( $\eta_\ell$=$e^{2i\delta_\ell}$=$\alpha_\ell e^{i\beta_\ell}$)  are  determined.  
For a given  set  of  potentials, these complex phase sifts are strongly angular momentum and energy dependent.
In terms  of  the  phase  shifts  and  the  wave  number ($ \lambdabar = \hbar
/\sqrt{2mE}$) and neglecting spins, various cross sections are given below. 
\begin{eqnarray}
\sigma_{tot}&=&2\pi\lambdabar^2\sum_\ell(2\ell+1)\left[1-\Re{\eta_\ell }\right] \label{st-om} \\
%\sigma_{se}&=&\pi\lambdabar^2\sum_\ell(2\ell +1)|1-\eta_\ell |^2  \label{se-om} \\
\sigma_{reac}&=&\pi\lambdabar^2 \sum_\ell(2\ell +1)\left(1-|\eta_\ell |^2\right)  \label{r-om} \\
\frac{d\sigma}{d\Omega}(\theta)&=&\frac{\lambdabar ^2}{4} \left|\sum_\ell (2\ell +1)(1-\eta_\ell )P_\ell (\cos{\theta})\right|^2  \label{s0-om}
\end{eqnarray}
In Figure \ref{npbtl}, we show the transmission as a function of angular momentum $\ell$ for n+$^{208}$Pb
system at  high and low energies. The results in this figure are generated using SCAT2 code \cite{scat2} with Koning and Deleroche potentials \cite{kd}.
The transmission, which is related to  absolute  value of the complex phase shift ($1-|\eta_\ell|^2$), 
is approximately constant up to some $\ell_{max}$ for each energy and falls sharply after that cut off 
angular momentum.. We have plotted the SCAT2 transmissions and seen that 
this angular momentum dependence is very similar in all the systems considered in this work. 
These results are well known in the field of heavy ion fusion around the Coulomb barrier energies 
where,  a sharp cut off  or diffuse cut-off models are often used. When multiplied by 
factor, The transmission functions give the well known triangular distribution shape of  partial wave cross sections 
( $\sigma_\ell=(\pi/k^2) (2\ell+1)$.  Ramsauer model in the following discussion is essentially based on this behaviour of transmission function. \\
\noindent
An excellent  experimental data  base  of  neutron total cross sections is presently
available in the energy range up to 600 MeV  \cite{finlay,dietrich1,abfal}.
The  most  recent  work  by Koning and Delaroche \cite{kd} presents a very
exhaustive search for OMP parameters that fit the data very well up to  200
MeV.  We made a phenomenological Ramsauer  model
analysis  of  the experimental total cross sections. The  nuclear
Ramsauer  model  \cite{lawson}  provides a simple method to parameterize the
energy dependence of neutron nucleus total scattering cross sections.  This
model  assumes  that the scattering phase shifts are independent of angular
momentum ($\ell$) as  reflected  in  Eq.\eqref{eqstfit}  ($\eta=\alpha  e^{i\beta})$,
in contrast  to the predictions of the optical model given in Eq.\eqref{st-om}. 
Further, it was proposed that the $\ell$-independent phase shift varies slowly  with
energy.  Peterson applied this model  for the study of  neutron  scattering
from  various  nuclei   \cite{peterson,book}.  There were some
attempts \cite{franco,gould,anderson,grimes1}  (see references therein)
to  put  this  Ramsauer  model  on  a  sound  theoretical   basis. The neutron
total cross sections have thus been well studied 
\cite{anderson,bauer,madsen,grimes1,grimes2,grimes3,dietrich2} using this model,  over  a
wide  range  of  nuclear  masses and neutron energies up to 500 MeV. 
In \cite{deb1,deb2},   simple functional forms
for the total cross sections through the $\ell_0$ parameter corresponding 
to  partial wave distributions were shown. In their method,
one can determine the partial wave cross sections,  considering  spin. 
\noindent
\section{Analysis of neutron total cross sections}
\begin{figure}
\includegraphics[width=9.00cm,height=16.0cm]{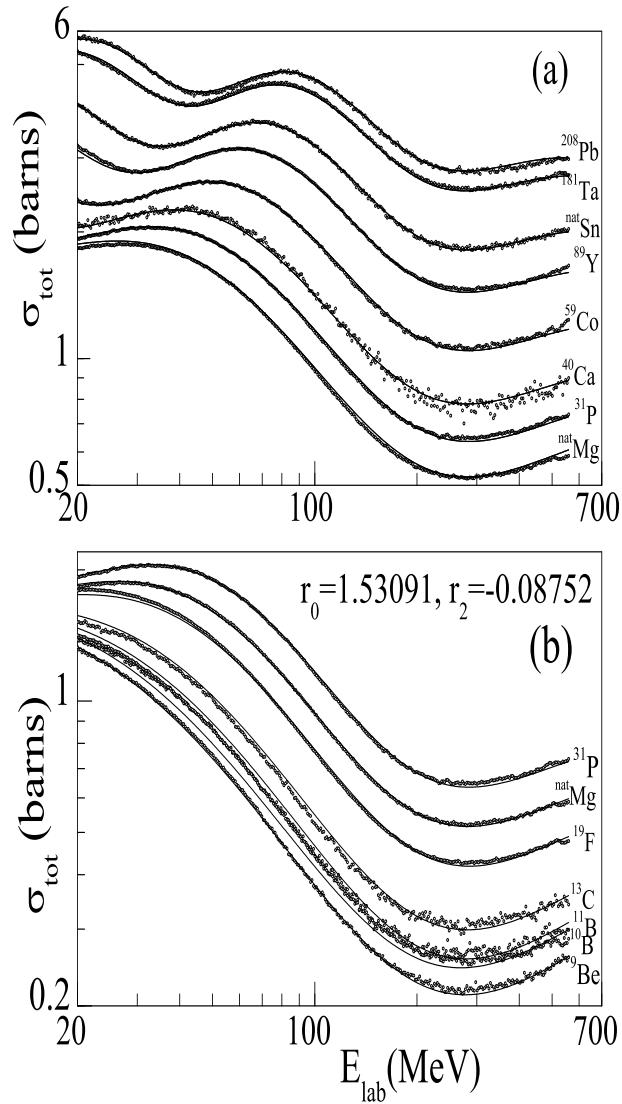}
\caption{Ramsauer  model  fits  (solid  lines)  to  the experimental data (symbols) 
of  total  cross sections   versus  E$_{lab}$  using  Eq.\eqref{eqstfit} \cite{surya1}.
The  twelve parameters optimised and the functional forms used are mentioned in the text in Eqs.\eqref{stfitfun}-\eqref{stfitpar}.
Fig. (a) is for heavy nuclei and Fig. (b) is for light nuclei. The radius parameters 
used for light systems are mentioned on the Figure (b).}
\label{sigtot}
\end{figure}

\begin{figure}
\includegraphics[width=9.25cm,height=16.0cm]{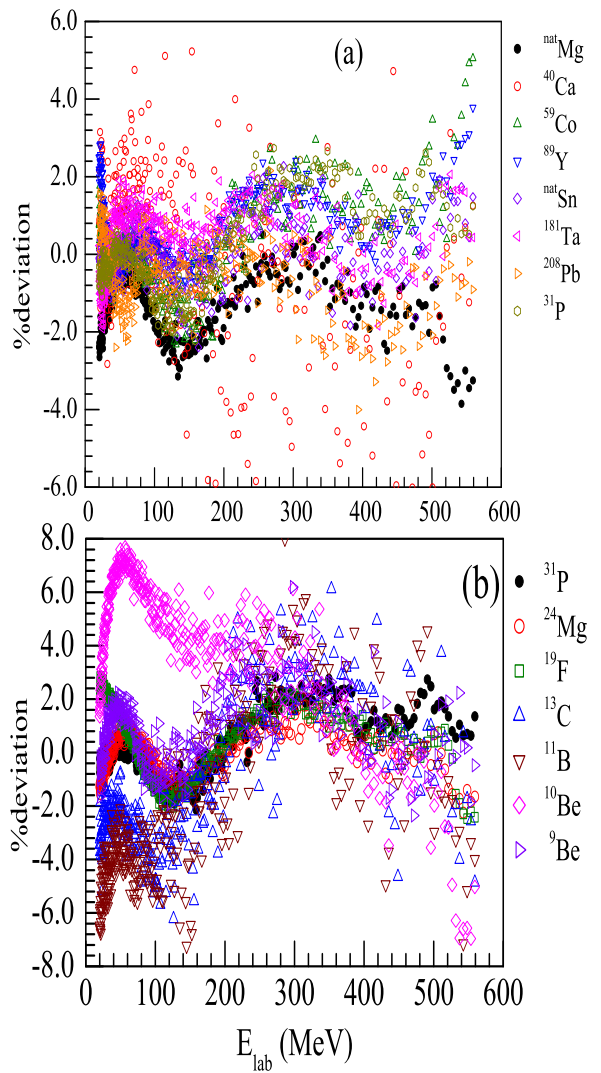}
\caption{The percentage deviations of the empirical fits in Figs. \ref{stglobal}(a,b) 
from the experimental data are shown for heavy and light nuclei respectively in Figs.(a,b).}
\label{devsigtot}
\end{figure}

\begin{figure}
\includegraphics[width=9.0cm,height=10.0cm]{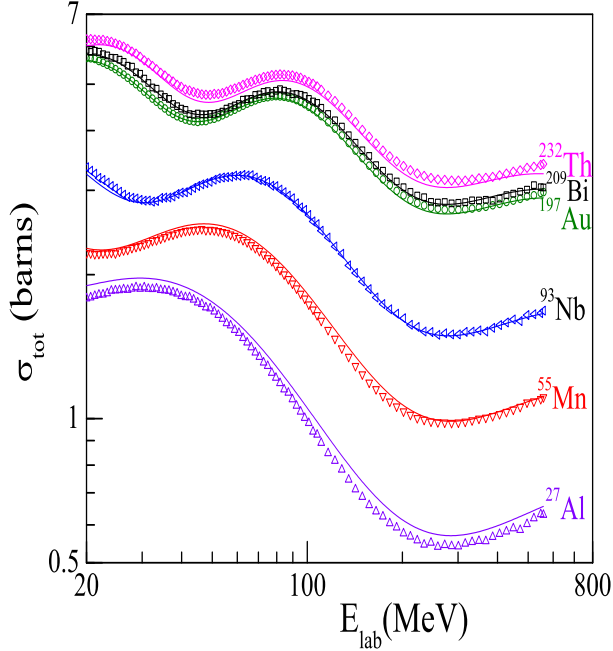}
\caption{The Ramsauer  model  predictions  (solid  lines)  to  the experimental data (symbols) 
of  total  cross sections   versus  E$_{lab}$  using  Eq.\eqref{eqstfit}, using the 
functions and parameters that were optimised for Figure. \ref{stglobal}(a) as in Eqs.\eqref{stfitfun}-\eqref{stfitpar}.}
\label{testheavy}
\end{figure}

\noindent
We  performed  the  Ramsauer  model  fits  to the experimental  neutron  total  cross 
sections using Eq.\eqref{eqstfit}.  The quantities $R (fm),\alpha$ and $\beta$  are functions 
of atomic mass number (A)  and the center of mass energy (E).
\begin{eqnarray}
\sigma_{tot}&=&{2\pi}\ (R+\lambdabar)^2\left(1-\alpha\cos{\beta}\right) \label{eqstfit} \\
%\end{eqnarray}
%\begin{eqnarray}
\beta&=&\beta_x A^{\frac{1}{3}}( \sqrt{E+V} - \sqrt{E} )\label{stfitfun}\\
V&=&V_A+V_E\sqrt{E} \nonumber \\
V_A&=&V_0+V_1 (N-Z)/A +V_2/A \nonumber\\
\alpha&=&\alpha_0+\alpha_A \sqrt{E} \label{st-alpa}\\
\alpha_A&=&\alpha_{1} \ln(A) +\alpha_{2}/\ln(A)  \nonumber\\
R&=&r_0 A^{\frac{1}{3}} + r_A \sqrt{E} + r_2  \label{st-rad}\\
r_A&=&r_{10} \ln(A) + r_{11}/\ln(A) \nonumber \\
r_0&=&1.42988,~~r_{10}=-0.02298,~~r_{11}=0.10268 \label{stfitpar}\\
r_2&=&0.23216,~~V_0=46.51099,~~V_1=6.73833 \nonumber\\
V_2&=&-117.52082,~~V_E=-3.21817,~~\beta_x=0.5928 \nonumber\\
\alpha_0&=&0.02868,~~\alpha_{1}=-0.00274,~~\alpha_{2}=0.13211 \nonumber
\end{eqnarray}
Figures \ref{sigtot}(a,b) show the experimental data represented by symbols and
 the Ramsauer model fits (solid lines) for $\sigma_{tot}$  cross  sections
using Eqs.\eqref{eqstfit}-\eqref{stfitpar}.
The fits are obtained with twelve optimised parameters as given in   Eq.\eqref{stfitpar} over  wide
mass  range of $^{24}$Mg to $^{208}$Pb (Fig. \ref{sigtot}(a))   and $^9$Be to $^{24}$Mg (Fig. \ref{sigtot}(b)) \cite{surya1}.
These fits   cover the neutron energy region ($E_{lab}$) of 20-550 MeV.
Our  functional  dependence  on  energy  and  mass given in Eq.\eqref{stfitfun}-\eqref{st-rad}
with twelve global parameters ( Eq.\eqref{stfitpar}) reproduced  the experimental
data  well \cite{surya1}. The percentage deviations of these fits from data  shown in Figs. \ref{sigtot}(a,b) 
have been calculated. We observed that the fits are within $\pm6\%$ for heavy nuclei
and within $\pm8\%$  for light nuclei  for energies between 20MeV-550MeV. 
These deviation functions are shown for all systems in Figs.\ref{devsigtot}(a,b).
As seen in Figures \ref{devsigtot},  overall fits are  within $8\%$ error from the data.  Note that the
radius parameters for light systems are different as mentioned on the Figure \ref{sigtot}(b).
In order to test these parameterization, we predicted $\sigma_{tot}$ using 
Eqs.\eqref{eqstfit}-\eqref{stfitpar} for six nuclei which were not included in global fits of Figure \ref{sigtot}. These predictions 
and the respective experimental data are shown in Figure \ref{testheavy}.  We have seen that the predictions are  within $8\%$ deviation from data.
Model fits to total cross sections were already shown by various groups  \cite{grimes1,grimes2,deb1,deb2}  (see  the  references  therein).Some works 
 reproduced very well the Ramsauer peak, superposed on their smoothly varying  parts of the cross sections \cite{deb2}.

%--------------figure 1----------------
%--------------figure 1----------------
\begin{figure}
\hoffset -1.50cm
\includegraphics[width=9.00cm,height=8.50cm]{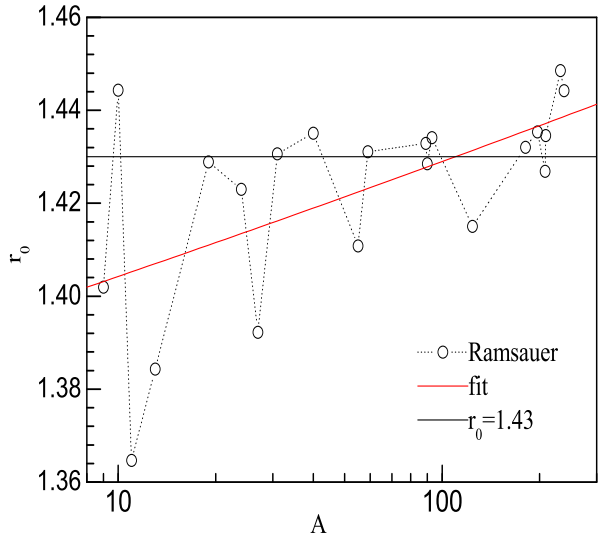}
\caption{The r$_0$ parameter values that result in best Ramsauer model description to experimental
data have been shown versus mass number.}
\label{r0param}
\end{figure}
\begin{figure}
\hoffset -1.5cm
\includegraphics[width=9.00cm,height=8.5cm]{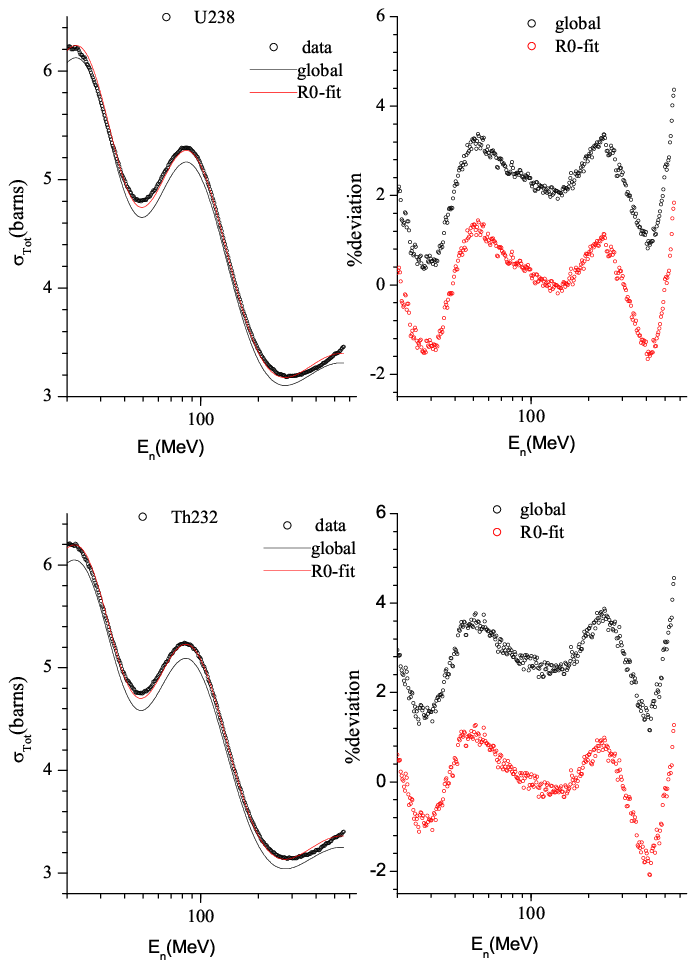}
\caption{Ramsauer  model  fits  (solid  lines)  to  the experimental data (symbols) 
of  total  cross sections   versus  E$_{lab}$  using  Eq.\eqref{eqstfit}. The global fits 
are same as in Fig.\ref{sigtot}. The red curve shows the fits by varying the $r_0$ parameter.
Left panel shows the fits to cross sections for $^{238}$U and $^{232}$Th systems and right 
panel shows the percentage deviation of respective fits from the data. The red symbols show 
less deviation compared to black symbols (global fits) as expected.}
\label{uth}
\end{figure}
\begin{figure}
\hoffset -1.50cm
\includegraphics[width=9.00cm,height=8.00cm]{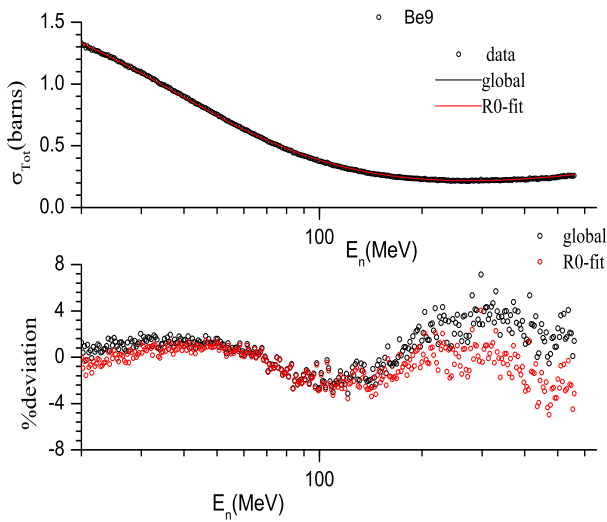}
\caption{As  shown in Figs.\ref{uth}, for $^{9}$Be.}
\label{be9}
\end{figure}

\noindent
The Ramsauer model global fits  of experimental neutron total  
cross sections are  shown in figures \ref{sigtot} and tested for systems which are not used in fitting process as shown in 
Fig.\ref{testheavy}. The fit functions  and global set of parameters are provided in Eqs.\eqref{stfitfun}-\eqref{st-rad}, \eqref{stfitpar}. 
These functions and parameters are a convenient starting point for Ramsauer model analysis of the experimental data 
and  further improved fits can be achieved by local finer variation of these parameters. 
In this context, we observed that by adjusting only a single parameter $r_0$, it is possible to achieve better fits to experimental
data, as  has been earlier shown by us for neutron total scattering on targets
$^{238}$U and $^{232}$Th \cite{insac}. We extended this study for all the systems, light and heavy  considered in 
figures \ref{sigtot},\ref{testheavy}. It is well known that the nuclear radius is given by r$_0$A$^{1/3}$.  For Ramsauer model 
analysis, this is a good approximation only for global fits. However, local minor 
deviations may occur for nuclear radii from this formula r$_0$A$^{1/3}$.  It is important to note  that this  radius, r$_0$A$^{1/3}$ ,
should not be taken to mean the actual radial size of the nucleus. It represents the energy independent part of the radial size required in the Ramsauer model
fits, see Eq.\eqref{st-rad}. Therefore, it represents the size required by neutron total cross sections, very similar to the strong absorption 
radius in heavy ion elastic scattering. Hence, this radial size is usually much larger than nuclear size. \\
Therefore we performed an elaborate study by  varying  the r$_0$
parameter and fixing  all other parameters as given in Eq.\ref{stfitpar}. We obtained  $r_0$ adjusted fits and their percentage 
deviations for all the  systems considered in this study and compared these with the global fits mentioned earlier.  
The required $r_0$ values are shown in Fig.\ref{r0param} and will be discussed later. These detailed fits and comparisons  are shown in 
Figs.\ref{uth},-\ref{be9}. In these figures, we show cross section fits in 
left panel and percentage deviation of fit from data in right panels. In these left panel figures, symbols represent data, black curves for global fits
and red curves for $r_0$ adjusted  fits.  In the right panels, we  show  $\%$deviations of global fits (black circles) and $r_0$ adjusted fits (red circles).
As can be seen in all these figures, the systematic trend in $\%$ deviations were removed by $r_0$ adjustment. The $\%$ deviations are in most cases 
within $\pm 4\%$.  It is owing to the requirement of 
a finer adjustment of $r_0$ parameter values, the global fits produced large  deviations up to $\pm8\%$. \\
As discussed earlier, the r$_0$ parameter values that result in best Ramsauer model description to experimental
data are shown in Fig.\ref{r0param}.  The constant value of $r_0=1.42988$ of global fits
is  shown in the figure by black straight line for reference.
The smooth red solid curve in Fig.\ref{r0param} shows the required variation of mass dependence of the $r_0$.
This is given by  a formula  $r_0(A)=1.0005+0.38045A^{0.02581}$.  This functional form and parameters show that $r_0A^{1/3}$
is an  approximation, however,  the  finer deviation of nuclear mass dependence  is shown by second term. 
The fine adjusted $r_0$ values using Ramsauer model are shown in Table.1. in column 2 and the from the formula given in column 3.
%The  Ramsauer model best fits results are respectively, $r_0=$1.4442,1.42689, 1.4311,1.4230 fm. 
These results in Table are to be compared with global fixed value of $r_0$=1.42990 . Using this dependence of $r_0$ values would 
result in  improved  Ramsauer model fits of the experimental data. 
\par
 %%------------------------------------------------
\begin{figure}
\hoffset -1.5cm
\includegraphics[width=9.00cm,height=8.50cm]{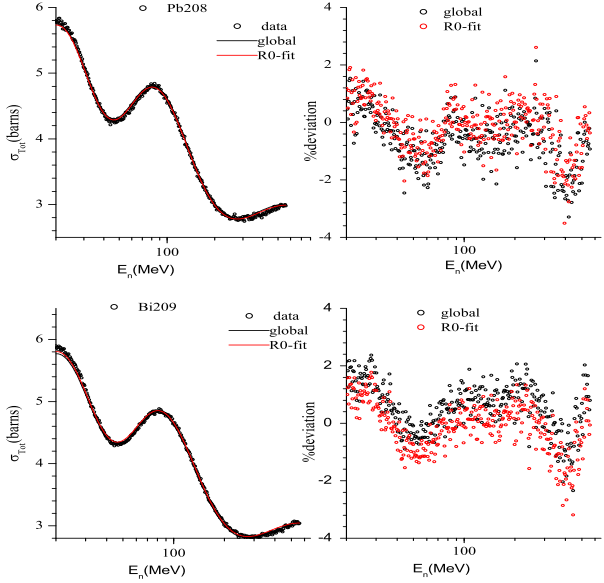}
\caption{As  shown in Figs.\ref{uth}, for $^{208}$Pb  and $^{209}$Bi.}
\label{pbbi}
\end{figure}
\begin{figure}
\hoffset -1.50cm
\includegraphics[width=9.00cm,height=8.50cm]{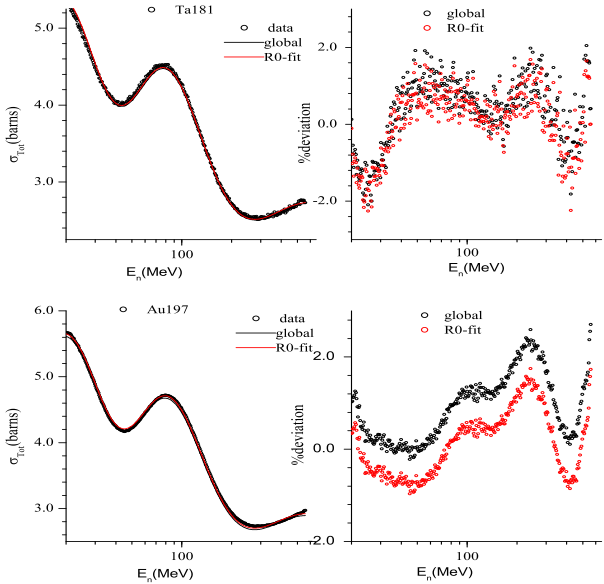}
\caption{As  shown in Figs.\ref{uth}, for $^{181}$Ta  and $^{197}$Au.}
\label{taau}
\end{figure}
\begin{figure}
\hoffset -1.5cm
\includegraphics[width=9.00cm,height=8.50cm]{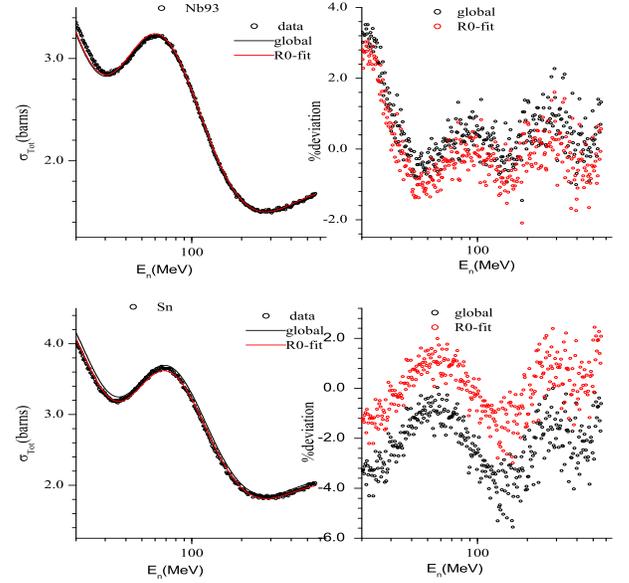}
\caption{As  shown in Figs.\ref{uth}, for $^{93}$Nb  and $^{nat}$Sn.}
\label{nbsn}
\end{figure}
\begin{figure}
\hoffset -1.5cm
\includegraphics[width=9.00cm,height=8.00cm]{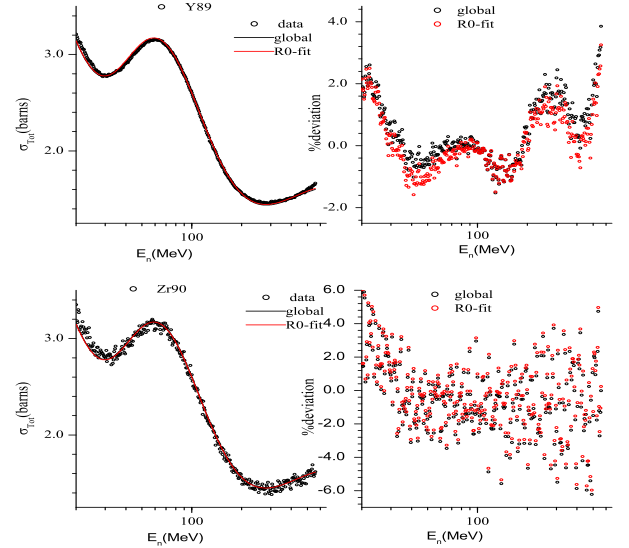}
\caption{As  shown in Figs.\ref{uth}, for $^{89}$Y  and $^{90}$Zr.}
\label{yzr}
\end{figure}
\begin{figure} \hoffset -1.5cm
\includegraphics[width=9.00cm,height=8.00cm]{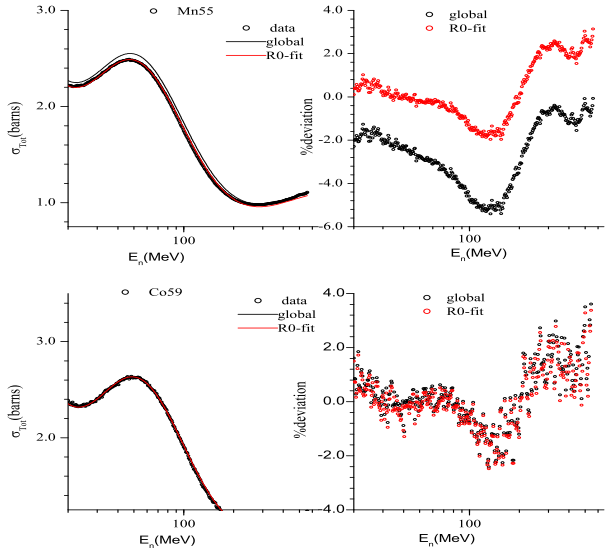}
\caption{As  shown in Figs.\ref{uth}, for $^{55}$Mn  and $^{59}$Co.}
\label{mnco}
\end{figure}
\begin{figure}
\hoffset -1.50cm
\includegraphics[width=9.00cm,height=8.00cm]{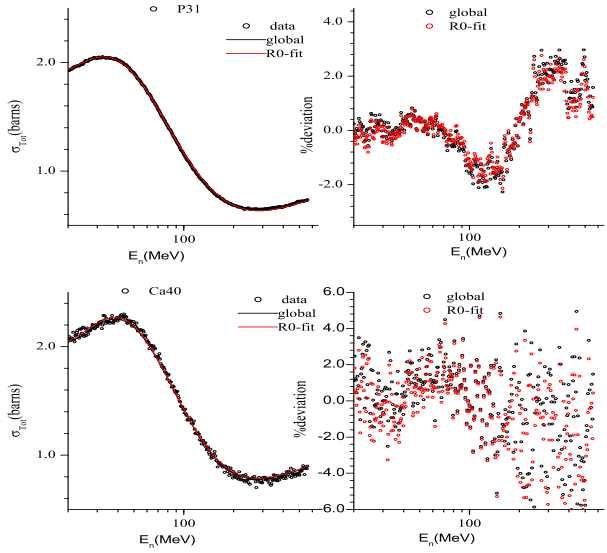}
\caption{As  shown in Figs.\ref{uth}, for $^{31}$P  and $^{40}$Ca.}
\label{pca}
\end{figure}
\begin{figure}
\hoffset -1.50cm
\includegraphics[width=9.00cm,height=8.00cm]{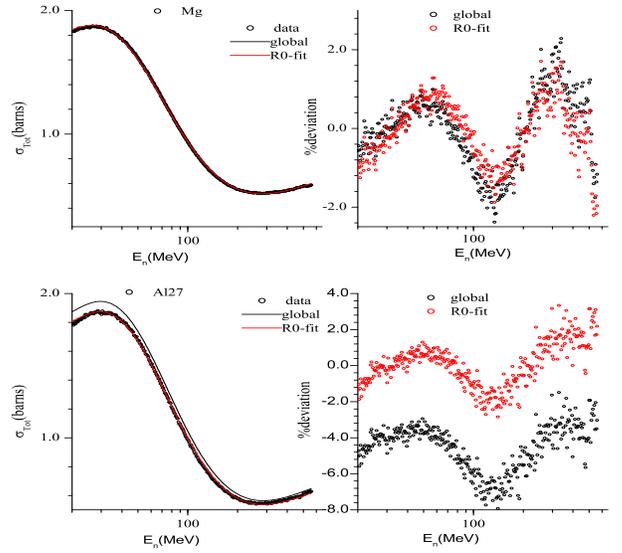}
\caption{As  shown in Figs.\ref{uth}, for $^{nat}$Mg  and $^{27}$Al.}
\label{mgal}
\end{figure}
\begin{figure}
\hoffset -1.50cm
\includegraphics[width=9.00cm,height=8.00cm]{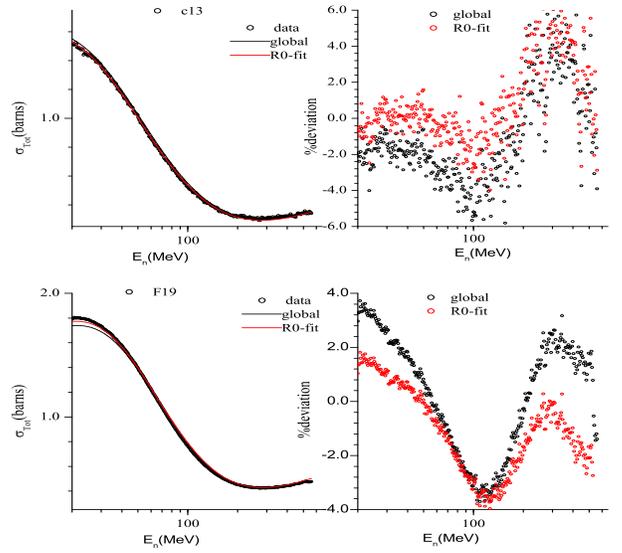}
\caption{As  shown in Figs.\ref{uth}, for $^{13}$C  and $^{19}$F.}
\label{c13f19}
\end{figure}
\begin{figure}
\hoffset -1.50cm
\includegraphics[width=9.00cm,height=8.00cm]{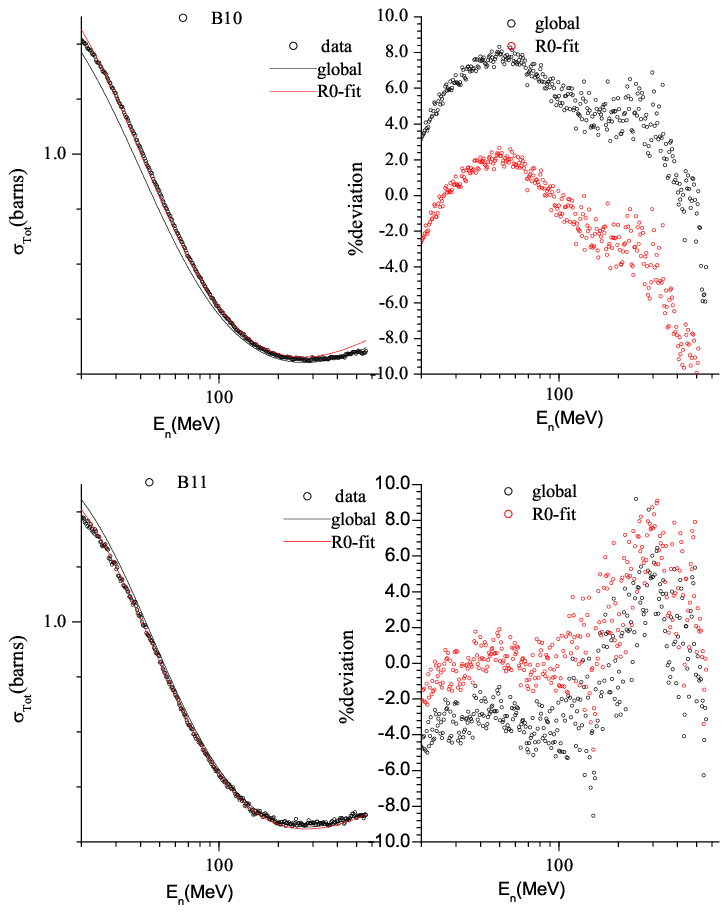}
\caption{As  shown in Figs.\ref{uth}, for $^{10,110}$B.}
\label{b10b11}
\end{figure}
\par
\begin{center}{
\begin{table}
\caption{Table shows $r_0$ values for various nuclei required to achieve
best fits using Ramsauer model for neutron total cross sections. 
First column shows nucleus, second column shows the $r_0$ value. This is to be compared 
with a fixed value of 1.43fm used in global analyses.}
\begin{ruledtabular}
\begin{tabular}{|c|c|c|}
\hline
Nucleus       	& $	r_0(fm)$ & formula      \\
\hline
$^{238}$U &	1.4442   &	1.43866  \\
$^{232}$Th & 1.44853 &  1.43838	 \\
$^{209}$Bi & 1.43457 & 1.43720 	 \\
$^{208}$Pb & 1.42689  & 1.43714	\\	
$^{197}$Au & 1.43534  & 1.43558	\\
$^{181}$Ta & 1.43204  & 1.43653	\\
$^{120}$Sn & 1.41503  & 1.43135	 \\
$^{93}$Nb & 1.43409   &	1.42817 	\\
$^{90}$Zr & 1.4285    & 1.42780	\\
$^{89}$Y & 1.43287    & 1.42768	 \\
$^{59}$Co & 1.43109   & 1.42317	 \\
$^{55}$Mn & 1.41083   & 1.42241	 \\
$^{40}$Ca & 1.43506   & 1.41895	 \\
$^{31}$P & 1.43062    & 1.41621	\\
$^{27}$Al & 1.39221   & 1.41473	\\
$^{24}$Mg & 1.42302   & 1.41347	 \\
$^{19}$F & 1.42892    & 1.41099	\\
$^{13}$C & 1.38432    & 1.40699	 \\
$^{11}$B & 1.36467    & 1.40524	\\
$^{10}$B & 1.44432    & 1.40425  	\\
$^{9}$Be & 1.40193    & 1.40315	\\
\hline
\end{tabular}
\end{ruledtabular}
\label{tabler0param}
\end{table}
}\end{center}
\par
\begin{center}{
\begin{table}
\caption{Table shows best adjusted $E_0,ae$ values of the f(E) function that fit the neutron total cross sections
for various nuclei using SCAT2 program. First column shows nucleus, second and third columns show the 
$E_c$,$E_0$ and a$_e$ values.}
\begin{ruledtabular}
\begin{tabular}{|c|c|c|c|}
\hline
Nucleus      & 	E$_c$(MeV) 	& 	E$_0$(MeV) & a$_e$ (MeV)     \\
\hline
$^{209}$Bi & 350 & 350 & 260  \\
$^{208}$Pb & 325 & 325 & 275  \\	
$^{197}$Au & 350 & 350 & 245	 \\
$^{181}$Ta & 300 & 300 & 280	 \\
$^{120}$Sn & 250 & 250 & 270	 \\
$^{93}$Nb  & 225 & 225 & 280  \\
$^{90}$Zr  & 225 & 225 & 280	 \\
$^{89}$Y   & 225 & 225 & 285  \\
$^{59}$Co  & 225 & 225 & 275	 \\
$^{55}$Mn  & 210 & 210 & 280	 \\
$^{40}$Ca  & 225 & 225 & 275	 \\
$^{31}$P   & 250 & 250 & 270  \\
$^{27}$Al  & 225 & 225 & 285  \\
$^{24}$Mg  & 250 & 250 & 275	 \\
$^{19}$F   & 250 & 250 & 275	 \\
$^{13}$C   & 250 & 250 & 300	 \\
\hline
\end{tabular}
\end{ruledtabular}
\label{kdtable}
\end{table}
}\end{center}
\noindent
\section {Optical potentials for neutron total cross sections}
\begin{figure}
\hoffset -1.50cm
\includegraphics[width=9.00cm,height=8.00cm]{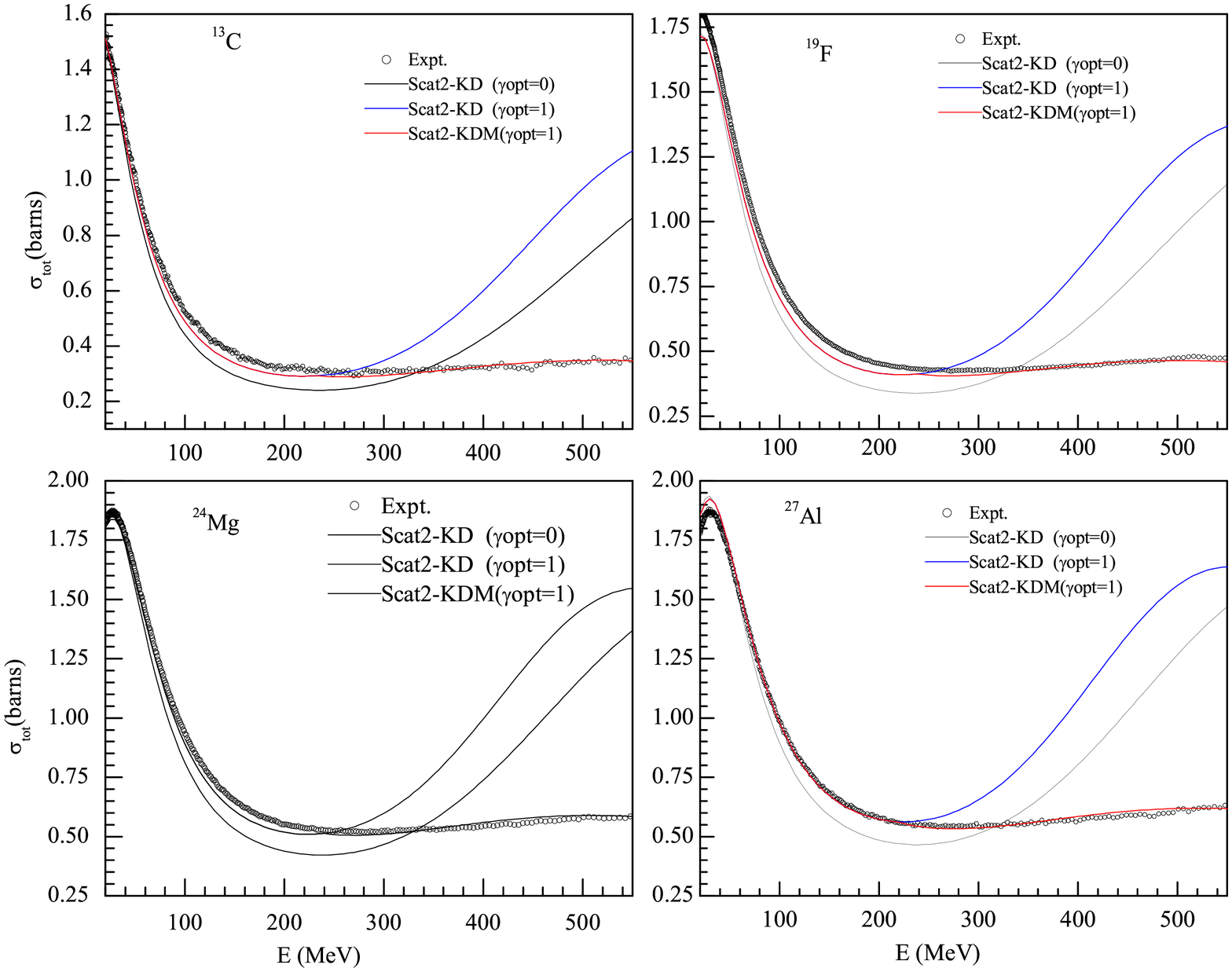}
\caption{Total cross sections for n scattering on $^{13}$C,$^{19}$F,$^{24}$Mg,$^{27}$Al. 
Symbols show experimental data. Black curve is SCAT2 results with KD potentaials with $\gamma=0$ option.
The blue curve represents with $\gamma=1$ option and red curves  represent modified KD 
potentials together with $\gamma=1$.}
\label{figcfmgal}
\end{figure}

\begin{figure}
\hoffset -1.50cm
\includegraphics[width=9.00cm,height=8.00cm]{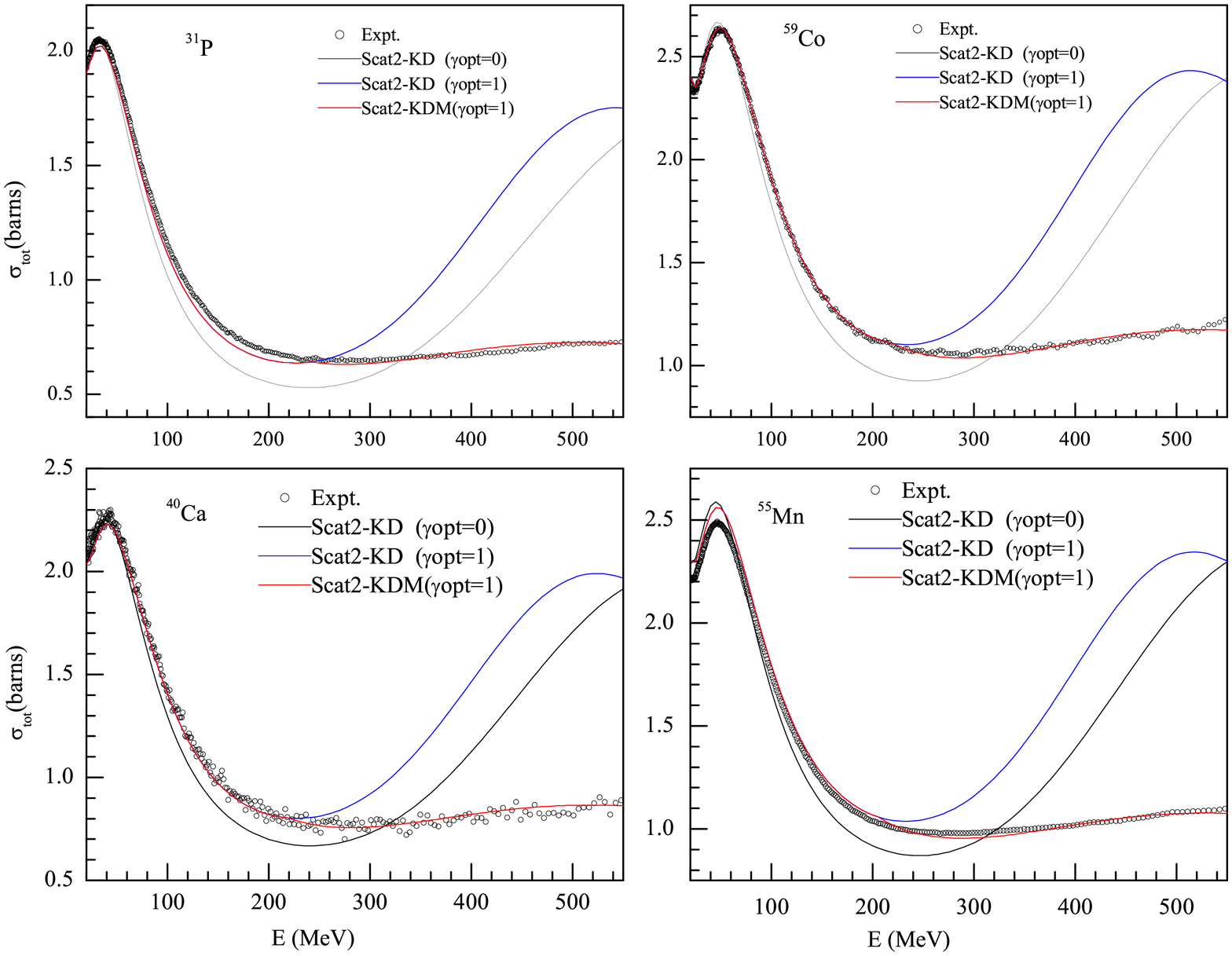}
\caption{Total cross sections for n scattering on $^{31}$P,$^{40}$Ca,$^{55}$Mn,$^{59}$Co. 
The  curves are as explained in previous figure.}
\label{figpcamnco}
\end{figure}

\begin{figure}
\hoffset -1.50cm
\includegraphics[width=9.00cm,height=8.00cm]{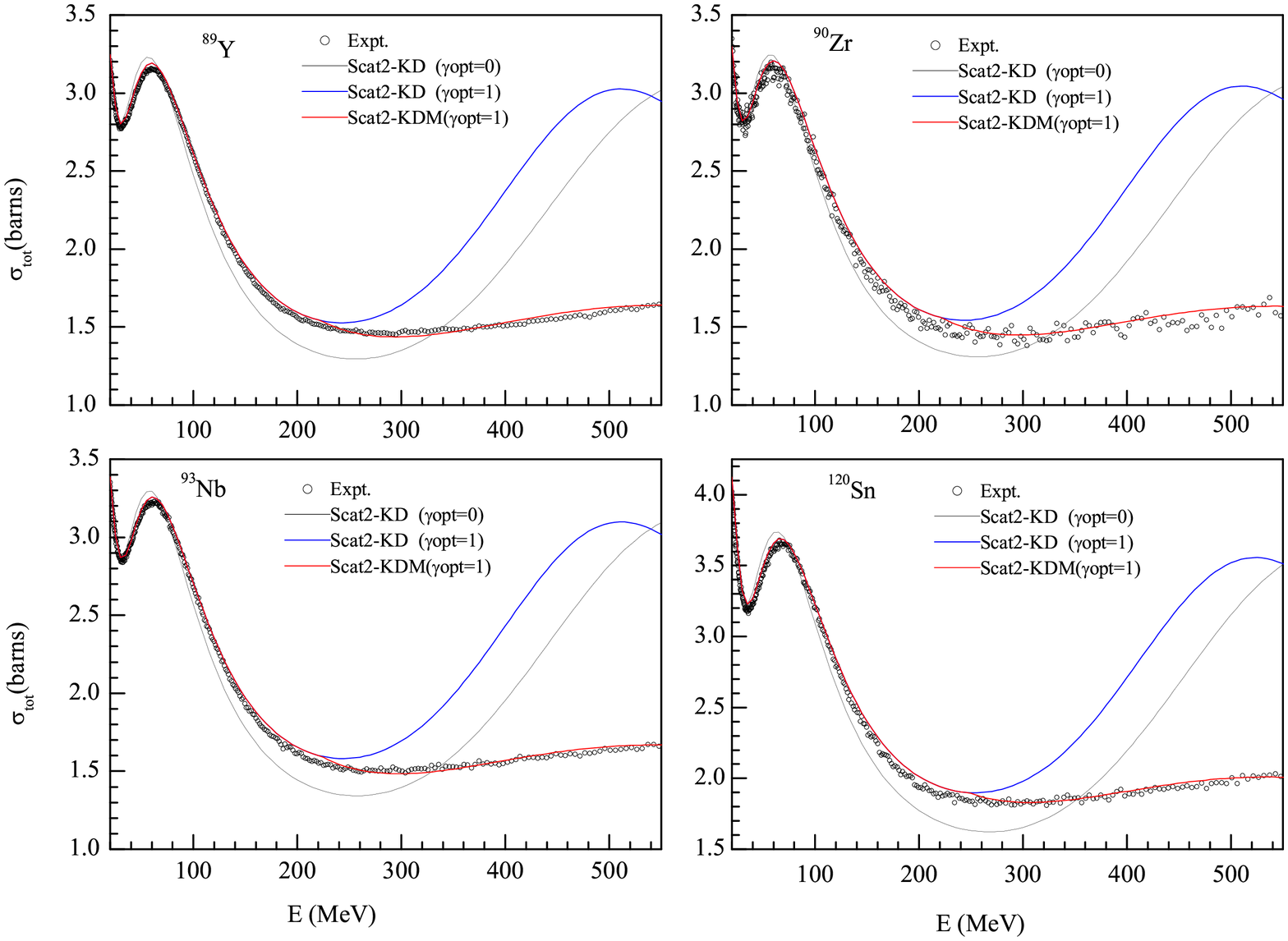}
\caption{Total cross sections for n scattering on $^{89}$Y,$^{90}$Zr,$^{93}$Nb,$^{120}$Sn. 
The  curves are as explained in previous figure.}
\label{figyzrnbsn}
\end{figure}

\begin{figure}
\hoffset -1.50cm
\includegraphics[width=9.00cm,height=8.00cm]{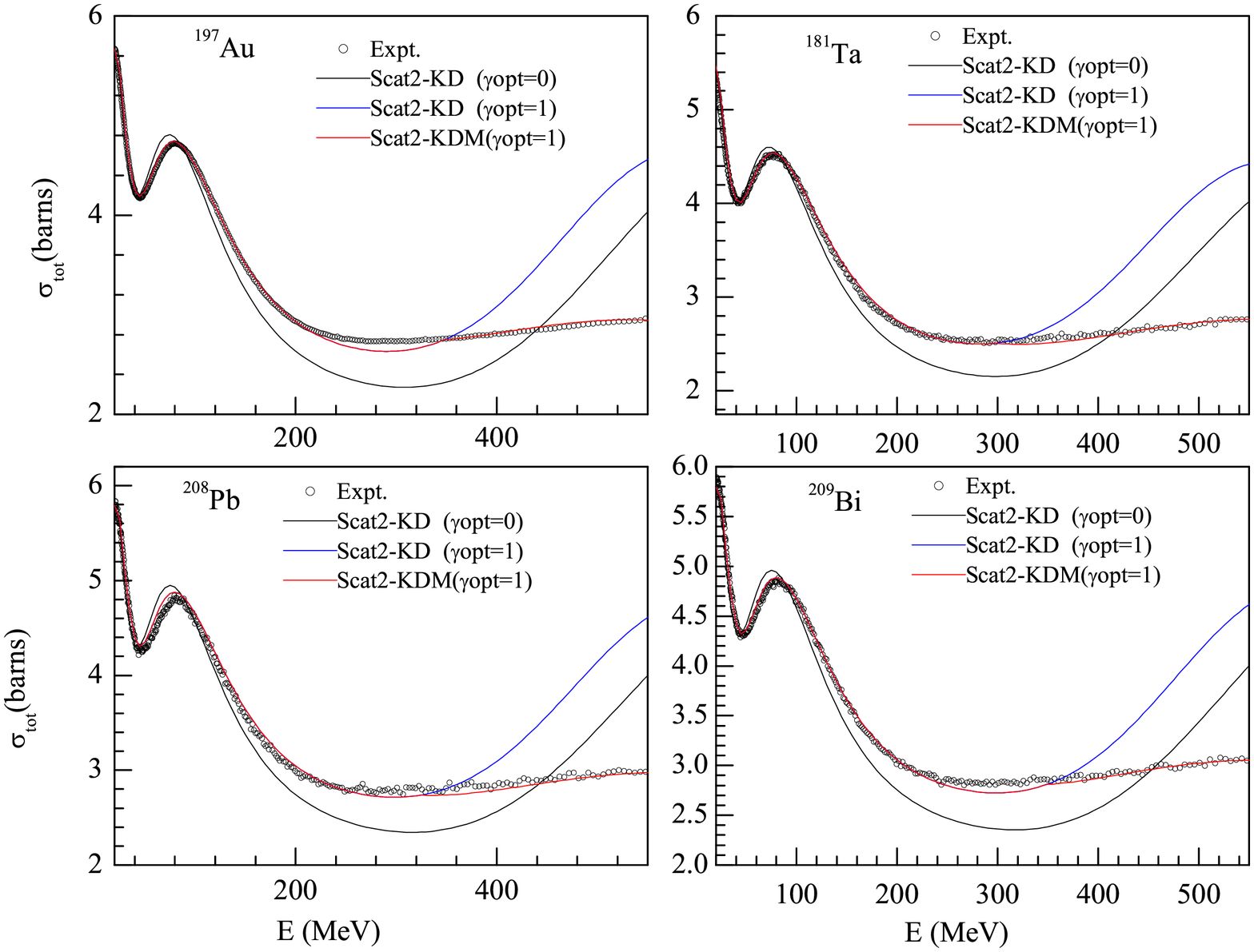}
\caption{Total cross sections for n scattering on $^{181}$Ta,$^{197}$Au,$^{208}$Pb,$^{209}$Bi. 
The  curves are as explained in previous figure.}
\label{figtaaupbbi}
\end{figure}

As mentioned in the introduction, the SCAT2 code \cite{scat2} with Koning Deleroche (KD) \cite{kd} 
potentials, explains consistently all the measured quantities of neutron nucleus cross sections,
such as total cross sections, angular distributions, shape elastic cross sections etc.
However, as well known its use is limited to 1KeV to 200 MeV energy region only. 
After achieving good fits using Ramsauer model, we have tried to 
modify the KD potentials in SCAT2 program such that they explain the total cross sections up to 550 MeV.
We use KD potentials up to energy E$_c$ and a modification function f(E) beyond E$_c$ for real potentials.
Effectively, f(E)=1 below $E_c$ and the f(E) function normalizes the  volume term of  real part of optical potentials, 
i.e., $\it pote(1)$ array in SCAT2 code. We have chosen exponential 
energy dependence factor given by $f(E)=\exp((E-E_0)/a_e)$, with
$E_0$ and ae and E$_c$ being adjusted for various nuclei and these values are shown in Table \ref{kdtable}.
In the following figures, we show these details with and without modification of KD potentials. 
In figures \ref{figcfmgal}-\ref{figtaaupbbi}, symbols show the experimental total cross sections of 
W. P. Abfalterer, F. B. Bateman, {\it et. al.,}\cite{abfal}. The black lines are for normal SCAT2 code 
results with the prescribed KD potentials with option $\gamma=0$. 
This option refers to no relativistic corrections in potentials.
THe blue curves represent  results with KD potentials with option $\gamma=1$.
As well known, this option refers to relativistic corrections implemented to 
optical potentials. The red curves show the total cross sections from our modified KD potentials using 
option $\gamma=1$. As seen in figures, the red curves simulate the neutron total cross sections well up to 550 MEV.  
\noindent
\section{Conclusion}
\noindent
In conclusion, we  performed  the Ramsauer  model  parameterization  of experimental 
neutron total  cross sections and derived the systematics of the Ramsauer  model parameters.
The percentage deviation of fits from data were plotted and displayed to be within $\pm8\%$. 
It was observed that the Ramsauer model fits are sensitive to nuclear radius parameter.
By fine adjustment of  $r_0$ parameter locally one can obtain very good
Ramsauer model fits to neutron total cross sections. The r$_0$  adjusted fits have been compared with the 
global fits for various systems. These  percentage deviations have  been presented in detail.
The r$_0$ parameter values that fit best have been shown as a function of nuclear mass. 
We modified the KD potentails incorporating new energy dependence. With these new potentials, the 
SCAT2 code was able to predict the total cross sections well up to 550 MeV.
\noindent
\acknowledgements
\noindent
We sincerely thank Dr. R.S. Gowda, whose references  cited below  were immensely useful in our follow-up work.
\par

%%%%%%--------------------------------------------------------------------

\begin{references}
\bibitem{ads1}  C.  Rubbia  {\it  et.  al.,}  " Conceptual design of a fast
Neutron Operated High power Energy Amplifier",
CERN Rep,   CERN/AT/95-44 (ET), Geneva, 29 Sep. 1995.
\bibitem{ads2}IAEA-TEC-DOC-985,  Accelerator  Driven Systems ; Energy generation
and Transmutation of Waste Status Report, Nov  1997,  International  Atomic
Energy Agency, Vienna.
\bibitem{lawson} J. D. Lawson, Phil. Mag., {\bf 44}, 102 (1953).
\bibitem{scat2} O. Bersillon, "SCAT2" Program, Note CEA-N-2227, Centre d'Etudes
Nucleaires de Bruyeres-Chatel, Service de Physique et techniques Nucleaires, France (Oct. 1981).
\bibitem{kd}A. J. Koning and J. -P. Delaroche, Nucl. Phys. {\bf A713}, 231 (2003).
\bibitem{finlay} R. W. Finlay, W. P. Abfalterer, G. Fink {\it et. al.,} Phys. Rev. {\bf C47}, 237 (1993).
\bibitem{dietrich1} F. S. Dietrich, W. P. Abfalterer, {\it et. al.,}
 $LANSCE-WNR$, Proc. Int. Conf. Nucl. data for Sci. and Tech.,
 Trieste, Italy, May 19-24, 1997, Vol.59, p.402, Italian Phys. Soc. (1997).
\bibitem{abfal}  W. P. Abfalterer, F. B. Bateman, {\it et. al.,} Phys. Rev {\bf C63}, 044608, (2001).
\bibitem{peterson} J. M. Peterson, Phys. Rev. {\bf 125}, 955 (1962).
\bibitem{book}  A.  Bohr and B. Mottelson, Nuclear Structure, Vol. 1 P.166,
Benjamin, N.Y. (1969).
\bibitem{franco} V. Franco, Phys. Rev. {\bf B140}, 1501 (1965).
\bibitem{gould} C. R. Gould {\it et. al.,} Phys. Rev. Lett. {\bf 53}, 2371 (1986).
\bibitem{anderson} J. D. Anderson and S. M. Grimes,
Phys. Rev. {\bf C41}, 2904 (1990).
\bibitem{grimes1} S. M. Grimes, J. D. Anderson, R. W. Bauer
and V. A. Madsen, Nucl. Sci. and Engg. {\bf 130}, 340 (1998).
\bibitem{madsen} V. A. Madsen, J. D. Anderson, S. M. Grimes, V. R. Brown and P. M. Antony,
Phys. Rev. {\bf C56}, 365 (1997).
\bibitem{bauer} R. W. Bauer {\it et. al.,} Nucl. Sci. and Engg. {\bf  130},
348 (1998).
\bibitem{grimes2} S. M. Grimes, J. D. Anderson, R. W. Bauer
and V. A. Madsen, Nucl. Sci. and Engg. {\bf 134}, 77 (2000).
\bibitem{grimes3} S. M. Grimes, J. D. Anderson and R. W. Bauer,
Nucl. Sci. and Engg. {\bf 135}, 296 (2000).
\bibitem{dietrich2} F. S. Dietrich, J. D. Anderson and R. W. Bauer, Phys. Rev. {\bf C68}, 064608 (2003).
\bibitem{deb1} P. K. Deb and K. Amos, Phys. Rev. {\bf 69}, 064608 (2004))~;~{\it ibid} {\bf 67}, 067602 (2003
\bibitem{deb2} P. K. Deb and K. Amos and S. Karataglidis Phys. Rev. {\bf 70}, 057601 (2004)~;~
P. K. Deb and K. Amos {\it. al., }Phys. Rev. Lett. {\bf 86}, 3248 (2001). 
\bibitem{gama01} M. Azam and R. S. Gowda, Nucl. Sci. and Engg. {\bf 144}, 1 (2003).
\bibitem{gama02} R. S. Gowda and S. Ganesan,   Nucl. Sci. and Engg, (2005).
%\bibitem{wick}G. C. Wick, Phys. Rev. {\bf 75}, 1459 (1949).
\bibitem{surya1}S. S. V. Suryanarayan, Rajesh S.Gowda and S. Ganesan, arXiv: nucl-th/0409005, 
{\it ibid.,}~nucl-th/0506004 
\bibitem{insac} S. V. Suryanarayan, S. Ganesan and R. S.Gowda, contributed paper T1-CP8, 
INSAC INPC conference held in Mumbai, India, 2005.
\end{references}
\end{document}